\documentclass[12pt,a4paper]{article}
\pdfoutput=1
\usepackage{amssymb,graphicx}
\usepackage{epstopdf}
\usepackage{epsfig}
\usepackage[utf8x]{inputenc}
\usepackage[english]{babel}
\usepackage{amsmath}
\usepackage{amsfonts}
\usepackage{amssymb}
\usepackage{fullpage}
\usepackage{ifpdf}
\usepackage{wrapfig}
\usepackage{booktabs}
\ifpdf
  \usepackage[pdftex,unicode]{hyperref}
  \hypersetup{
    pdftitle={Vector bosons escaping from the brane},
    pdfauthor={D.I.~Astakhov,D.V.~Kirpichnikov},
    colorlinks=true,
    citecolor=green,
    linkcolor=blue,
    urlcolor =blue,
    bookmarks=true,
    pdfpagemode=None
  }
\else
  \usepackage{hyperref}
\fi

\def\d{\partial}

\newcommand{\be}{\begin{equation}}
\newcommand{\ee}{\end{equation}}
\newcommand{\bea}{\begin{eqnarray}}
\newcommand{\eea}{\end{eqnarray}}
\newcommand{\bg}{\begin{gather}}
\newcommand{\eg}{\end{gather}}
\newcommand{\bseq}{\begin{subequations}}
\newcommand{\eseq}{\end{subequations}}
\newcommand{\tg}{\mathop{\rm tg}\nolimits}

\graphicspath{{images/}}

\begin{document}
\begin{flushright}
\end{flushright}
\vspace{10pt}
\begin{center}
  {\LARGE \bf Vector bosons escaping from the brane:\\[0.3cm]
$e^+e^- \to \gamma + \mbox{nothing}$ } \\
\vspace{20pt}
D.I.~Astakhov$^{a,b}$, D.V.~Kirpichnikov$^{a,c}$ 

\vspace{15pt}
\textit{$^a$Institute for Nuclear Research of the Russian Academy of Sciences,\\  60th October Anniversary
  Prospect, 7a, 117312 Moscow, Russia}

\textit{$^b$Moscow Institute of Physics and Technology,\\ Institutskii per. 9,  Dolgoprudny, 141700 Moscow Region, Russia}

\textit{$^c$Moscow State University, Department of Physics, \\
Vorobjevy Gory, 119991 Moscow, Russia}
    \end{center}
    \vspace{5pt}

\begin{abstract}
We consider phenomenological consequences of RS2-$n$
model with infinite extra dimension, namely,
the production, in high energy collisions, of a  photon 
 or $Z$-boson escaping from our brane in association with a photon 
remaining on the brane. This would show up as process
$e^+e^-\rightarrow \gamma + \mbox{nothing}$. We compare the
 signal with the
Standard Model background
coming from $e^+e^- \rightarrow \gamma \bar{\nu} \nu$.
We also make a comparison with $e^+e^-\rightarrow \gamma + \mbox{nothing}$
in the ADD model.
\end{abstract}
\section{Introduction}

Particle escape from our brane is a rather common feature
of brane-world theories with extra dimensions of infinite size.
This property has been already discussed  in the context of
the early toy models of brane world~\cite{Rubakov:1983bb}.
In $D$-brane constructions, incomplete localization of
matter and gauge fields on the brane, and hence particle escape from
the brane, is a characteristic of the Higgs phase~\cite{Dubovsky:2002bv}.
Warped models~\cite{Randall:1999vf} with the gravitational mechanism of
particle (quasi-)localization on the brane~\cite{Bajc:1999mh,Oda:2000zc}
also share the property of particle
escape~\cite{Dubovsky:2000am,Dubovsky:2000av}.
From 4-dimensional perspective,
particle escape occurs when there exists a continuum of massive
modes; in multi-dimensional language, these modes correspond to particles
moving away from the brane towards infinity in extra dimensions.
Even though particle escape can be interpreted in purely
4-dimensional terms via adS/CFT
correspondence~\cite{Gregory:2000rh,Giddings:2000ay}, in practice the escaped
particles are undetectable.   Their production
would manifest itself as missing energy events at particle
colliders.

In this paper we focus on photon and $Z$-boson escape from our brane
in a model with one non-compact warped extra dimension. This model
extends the original Randall--Sundrum brane-world set up~\cite{Randall:1999vf}
by the addition of $n$ compact extra
dimensions~\cite{Gregory:1999gv,Gherghetta:2000qi,Oda:2000zc,Dubovsky:2000av} (RS2-$n$ model). \pagebreak
 The latter ones are instrumental for the gravitational mechanism
of (quasi-)localization of gauge fields, similar to the graviton
localization in the original RS2 model. Specifically, the set up involves
a $(3 + n)$ --- brane with $n$ compact dimensions
 and positive tension, embedded in a $(5+n)$-dimensional space-time with
the $AdS_{5+n}$ metric:
\begin{equation}
  ds^2
=a(z)^2(\eta_{\mu \nu} dx^\mu dx^\nu - \delta_{ij} d \theta^i d \theta^j) -dz^2
\; .
  \label{gauge_metric}
\end{equation}
Here $\theta_{i}$ are the coordinates of
the compact extra-dimensions,
$\theta_{i}\in[0, 2\pi R_i]$,  $i=1, \dots n$,
and
\[
a(z)=e^{-k|z|}
\]
 is the warp factor characteristic of the Randall-Sundrum
class of models. This metric is a solution to the $(5+n)$-dimensional
 Einstein equations with the appropriately tuned brane tension.
The  $AdS_{5+n}$ curvature parameter $k$ is determined by
the $(5+n)$-dimensional Planck mass and the negative bulk cosmological constant.

In the background geometry (\ref{gauge_metric}), massless gauge field propagating in the bulk has an exactly localized
mode with zero 4-dimensional mass~\cite{Oda:2000zc,Dubovsky:2000av}.
The wave function of this mode is independent of extra-dimensional coordinates,
in accord with the requirement of charge universality~\cite{Dubovsky:2001pe}:
the 4-dimensional gauge charges of particles trapped to our brane
are independent of the shapes of their wave functions in extra dimensions.
On the other hand, once the gauge field obtains a mass via the Higgs mechanism
(with either brane or bulk Higgs field), this would-be localized mode
becomes quasi-localized on the brane, i.e., the massive vector boson
has finite width against the escape into extra dimensions.
In both cases, the continuum of the bulk modes starts from zero
4-dimensional mass squared.

There are both low- and high-energy manifestations of this scenario.
The low-energy phenomena include invisible positronium
decay~\cite{positronium}, star cooling~\cite{starcooling}, etc.,
and, indeed, strong constraints on the parameter $k$ have been
obtained, especially for small $n$, by considering these effects.
These low-energy effects may not be generic, however, since
in constructions generalizing \eqref{gauge_metric},
there might be a gap in the spectrum of 4-dimensional masses and/or
the bulk vector boson wave functions might be strongly suppressed on
the brane at low energies. One of the high energy manifestations
is the invisible $Z$-boson decay~\cite{Zdecay}; we quote the
corresponding constraints later on. In this paper we study yet another
effect, $e^+e^- \to \gamma +
\mbox{nothing}$, where ``nothing'' is an undetectable
photon or $Z$-boson escaping into extra dimensions. Our purpose is
to calculate the cross sections of this process in two
versions of the model and
figure out whether there are ranges of the parameter $k$ for different $n$
which would be accessible at future colliders.

The main Standard Model background to the process we discuss
is $e^+ e^- \to \gamma \bar{\nu} \nu$. So, we compare our signal
with this background. We also make comparison with the
ADD model~\cite{ArkaniHamed:1998rs} in which the
emission of a Kaluza--Klein graviton also shows up as the
process
$e^+e^- \to \gamma + \mbox{nothing}$, which  may also occur at sizeable
cross section~\cite{Giudice:1998ck}.

This paper is organized as follows. We begin in Section~\ref{2}
with a prototype model of one massless or massive bulk vector field
propagating in the background~\eqref{gauge_metric}. We assume everywhere
in this paper that fermions interacting
with the vector fields are confined to our brane and consider
in Section~\ref{2} vector-like interaction. We obtain the
wave functions of the bulk vector bosons, evaluate the 4-dimensional
effective action and arrive at the cross section of
$e^+e^- \to \gamma + \mbox{nothing}$ in this prototype model.
We study two versions of the Standard Model
embedded into the higher-dimensional theory in Sections~\ref{3}
and \ref{4}, respectively. In Section~\ref{3} we put   entire $SU(2)_W \times U(1)_Y$
gauge sector, as well as the Higgs sector, into the bulk.
This version is a straightforward generalization of the model
of Section~\ref{2}.
In Section~\ref{4} we assume that only $U(1)_Y$ part of
the Standard Model gauge theory lives in the bulk, and that
the Higgs field is localized on the brane. Although 
physics is somewhat different in this model, the results are
qualitatively similar to those of Section~\ref{3}.
In Section~\ref{section:signals} we derive the bounds on the parameter $k$ for various $n$, which are
based on the measurement of the invisible $Z$-decay width,
and  give comparison of $e^+e^- \to \gamma + \mbox{nothing}$
in the two models with the Standard Model  $e^+ e^- \to \gamma \bar{\nu} \nu$
and the ADD  process $e^+e^- \to \gamma + \mbox{KK~graviton}$.
We conclude in Section~\ref{section:conclusion}.

\section{Prototype model}
\label{2}

\subsection{Wave functions in extra dimensions}
\label{section:wave_functions}
We begin with a model of one bulk vector field $B_A$ coupled
to fermions localized on the brane.
The action in the background  metric~(\ref{gauge_metric})
 has the following form:
\begin{equation}
     S=\int \prod^n_{i=1} \dfrac{d\theta_i}{2\pi R_i}\,dz\, d^4x \sqrt{|g|}\,\left(-\frac{1}{4} g^{AC} g^{BD} F_{AB} F_{CD} + \frac{M_5^2}{2}g^{AB}B_{A} B_{B} \right) \label{eqn:vector_action}
\end{equation}
where the indices $A,B,C,D$ run from
$0$ to $(3+1+n)$, and $F_{CD}=\partial_{C} B_{D} - \partial_{D} B_{C}$.
The interaction
of the vector field with fermions is described by action:
\begin{equation}
  S_{int}=g_{5}\int\prod^n_{i=1} \dfrac{d\theta_i}{2\pi R_i}\,dz\,
d^4x \sqrt{g}\, \delta(z) \bar{\psi}(x) \gamma^\mu B_\mu(x,z) \psi(x)
  \label{eqn:gauge_interaction}
\end{equation}
where the
index $\mu$ runs
from $0$ to $3$ and $g_{5}$ is the
coupling constant of dimension $m^{-1/2}$.

 We consider the case where the
energy of colliding particles is smaller than $k$, and assume
that $R_i \lesssim k^{-1}$.
Thus, at low energies, $E\ll 1/R_{i}$, the relevant vector field modes
are independent of $\theta_{i}$.
We integrate the
action (\ref{eqn:vector_action}), (\ref{eqn:gauge_interaction})
over coordinates of the compact
extra dimensions and obtain the effective 5-dimensional action for these
modes:
\begin{multline}
     S=
\int  d^4x\, dz\, a^{n+4} \,\left(-\frac{1}{4} g^{ac} g^{bd} F_{ab} F_{cd} 
+ \frac{M_5^2}{2} g^{ab} B_a B_b \right) + \\
    + g_{5}\int d^4 x\, dz\, \delta(z)\,
\bar{\psi}(x) \gamma^\mu B_\mu(x,z) \psi(x)
     \label{formula:gauge_action_int}
\end{multline}
where the indices $a,b,c,d$ run from $0$ to $(3+1)$.
We now solve the classical equations of motion for the vector field
 in the background metric (\ref{gauge_metric}),
which follow from action~(\ref{formula:gauge_action_int}):
\begin{equation}
    \frac{1}{\sqrt{|g|}}\partial_c(\sqrt{|g|}g^{ac}g^{bd}F_{ab})+M_5^2 B^d=0
\; .  \label{eqn:vector_field}
\end{equation}
We do that  separately for massless and massive bulk vector fields.

\subsubsection{Massless bulk vector field}
\label{subsubsec:massless}
First, we consider the case of zero bulk mass, $M_5 = 0$, 
and reproduce the results of Ref.~\cite{Dubovsky:2000av}
(for a review see, e.g., Ref.~\cite{Rubakov:2001kp}).
Let us fix the gauge $B_z=0$. Then
Eq.~(\ref{eqn:vector_field}) reduces to two equations,
\begin{equation}
    \frac{1}{a^2}\partial_\mu F_{\mu \nu} - \frac{1}{a^{n+2}}\partial_z(a^{n+2}\partial_z B_\nu)=0 \label{eqn:maxvell_analog}
\end{equation}
\begin{equation}
    \partial_\mu \partial_z B_\mu = 0
\end{equation}
These equations describe a localized mode and a continuous spectrum of
massive excitations. The localized mode obeys
\begin{equation}
    \partial_z B_\mu=0 \; ,
    \label{eqn:photon_solution}
\end{equation}
so Eq.~(\ref{eqn:maxvell_analog}) becomes the
 Maxwell equation for free electromagnetic field.
Thus, the wave function of the localized mode is independent
of $z$. Nevertheless, it is normalizable, since the
appropriate weight is
$a^n(z) dz$, see \eqref{formula:gauge_action_int}.
We refer to this solution as $A_\mu$ in order to distinguish it from  non-localized ones discussed below.

The second type of solutions
corresponds to the continuous spectrum of
excitations, which are not localized on the
brane. From the 4-dimensional viewpoint  they have non-zero masses.
In the 4-dimensional momentum representation, the solutions
even under reflection of $z$
are
\be
  B_\mu(p,z)=B_\mu(p,m)  \Psi(z,m) \; ,
\label{gauge_solution}
\ee
with
\be
  \Psi(z,m) = C_m e^{(\frac{n}{2}+1)k|z|}\left(\eta_m J_{\frac{n}{2}+1}
\left(\frac{m}{k}e^{k|z|}\right)+N_{\frac{n}{2}+1}
\left(\frac{m}{k}e^{k|z|}\right)\right)
\label{formula:vertex_prefactor}
\ee
and
\be
  \eta_m =  -
\dfrac{N_{\frac{n}{2}}(\frac{m}{k})}{J_{\frac{n}{2}}(\frac{m}{k})}\; ,
\label{chi_m}
\ee
where $p^2=m^2$, $B_\mu(p,m)$  is transverse in 4-dimensional sense,
 $p^\mu B_{\mu}(p) = 0$, and  $C_m$
is the normalization constant.
Other possible modes, which are odd under reflection of $z$, do
not interact with fermions, and thus are of no interest.

One way to obtain the normalization condition is to consider the energy integral,
\begin{equation}
      E = \int d^4x\, dz \, T^0_0 \sqrt{|g|}
      \label{normalisation:energy}
\end{equation}
where $T^0_0$ is the energy-momentum tensor derived from the
action~(\ref{eqn:vector_action}). We substitute here the expansion of the
field~$B_\mu(x,z)$ in creation and annihilation operators,
\begin{equation}
    B_\mu(x,z)=\int_0^{\infty} dm \int \frac{d^3p}{(2\pi)^3} \cdot \frac{\Psi(z,m)}{\sqrt{2 E_{p,m}}}\sum_\alpha \left(a^{\alpha +}_{p,m} e^{ipx} + a^{\alpha }_{p,m} e^{-ipx}  \right)e^\alpha_\mu(p)\,,
\end{equation}
and require that the energy has the standard form
\begin{equation}
    E = \int_0^\infty dm \int d^3p\, E_{p,m}\sum_\alpha a^{\alpha+}_{p,m}a^{\alpha}_{p,m}\,.
\end{equation}
In this way we obtain the following normalization condition:
\begin{equation}
    \int dz a^{n}  \Psi(z,m) \Psi(z,m') =\delta(m-m') \; ,
    \label{eqn:gauge_normalisation}
\end{equation}
which gives
\begin{equation}
    C_m=\sqrt{ \frac{m}{2k(\eta_m^2+1)} } \; .
    \label{formula:normalisation_const}
\end{equation}
Although the normalization condition (\ref{eqn:gauge_normalisation}) is obvious,
the above way to derive it is quite general, and will be used 
below 
in less trivial situations.

By performing the integration in
(\ref{eqn:vector_action}) over the
coordinate of extra dimension $z$ and taking into account  the normalization
condition (\ref{eqn:gauge_normalisation}), we obtain the
following expression for the effective 4-dimensional
action  in the case  $M_5=0$:
\begin{multline}
	S_{eff}=-\frac{1}{4}
  \int d^4x\, \mathcal{F}_{\mu\nu}\mathcal{F}_{\mu\nu} +g_4 \int d^4x \, A_\mu \bar{\psi} 
\gamma_\mu \psi+ \\
+ \int dm\, d^4x \left(-\frac{1}{4} F_{\mu\nu}(x,m)F_{\mu\nu}(x,m)
+ \frac{m^2}{2}B_\mu(x,m)B_\mu(x,m) \right)+ \\
  +g_5 \int dm\, \Psi(0,m) \int d^4x\, 
B_\mu(x,m) \bar{\psi}\gamma^\mu\psi
	\label{formula:effective_action}
\end{multline}
The first two terms here
represent massless vector field
(\ref{eqn:photon_solution})  localized on the brane.
The effective 4-dimensional and 5-dimensional
couplings are related by
\be
g_4 = g_5 \cdot \sqrt{\frac{kn}{2}} \; ,
\label{eq:aug18-2}
\ee
where the factor
$\sqrt{\frac{kn}{2}}$ emerges due to the integration over $z$.
The interaction of
massive modes in
(\ref{formula:effective_action})  is suppressed at $m\ll k$ by
their wave functions at the brane position,
\begin{equation}
\Psi(0,m)= \frac{2}{ n \Gamma(\frac{n}{2}) }\cdot
\left( \frac{m}{2k}\right)^{\frac{n-1}{2}}\;
\label{A0}
\end{equation}
It follows from \eqref{formula:effective_action} that the phase space volume
element for modes from the continuous spectrum is
 \begin{equation}
    d\omega =
\dfrac{1}{2E_{{\bf{p}},m}}\dfrac{d^3p}{(2\pi)^3} d m, \qquad E_{{\bf{p}},m}
=\sqrt{m^2+{\bf{p}}^2}
	\label{formula:5D_phase_volume}
\end{equation}

\subsubsection{Massive field}

Now let us turn to the massive bulk vector field,
and focus on the case
\[
M_5 \ll k \; . 
\]
Equation
(\ref{eqn:vector_field}) now gives the following two equations:
\begin{multline}
    \frac{1}{a^2}(\partial^2_\mu B_{\nu}
-\partial_\nu\partial_\mu B_\mu))+k(n+2)
\mbox{sign}(z)\partial_z B_\mu -\partial^2_z B_\nu + M_5^2 B_\nu = \\
  = k(n+2)
\mbox{sign}(z)
\partial_\nu B_z - \partial_\nu\partial_z B_z
\label{eqn:massive_vector}
\end{multline}
\begin{equation}
    \partial_z \partial_\mu B_\mu =  \partial_\mu \d_\mu
B_z + M_5^2 a^2(z) B_z 
\label{equation for z comp}
\end{equation}
These  equations have transverse and longitudinal solutions in
4-dimensional sense.
Transverse solutions
obeying  $\partial_\mu B_\mu =0 $ have the following form:
\be
    B_\mu (p,z) = B_\mu (p,m)\, \Psi(z,m) \; , \;\;\;\;\; B_z=0 \; ,
\nonumber
\ee
where $p^2=m^2$,
\begin{eqnarray}
    \Psi(z,m)& = & \sqrt{ \frac{m}{2k(\chi_m^2+1)} } 
                \cdot e^{k(\frac{n}{2}+1)|z|}\left(\chi_m J_\nu
                \left(\frac{m}{k}e^{k|z|} \right)
                + N_\nu \left(\frac{m}{k}e^{k|z|} \right)  \right)
    \label{eq:aug18-1}
    \\
    \nu&=&\sqrt{\left(\frac{n}{2}+1 \right)^2+ \frac{M_5^2}{k^2}}
    \label{formula:massive_nu} \\
    \chi_m& =& - \frac{(\frac{n}{2}+1 -\nu)N_{\nu}(\frac{m}{k}) + \frac{m}{k} N_{\nu - 1}(\frac{m}{k})}
                      {(\frac{n}{2}+1 -\nu)J_{\nu}(\frac{m}{k}) + \frac{m}{k} J_{\nu - 1}(\frac{m}{k})}
\end{eqnarray}
Here we have used the
normalization condition (\ref{eqn:gauge_normalisation}).
For completeness, we also present the longitudinal solution:
\begin{eqnarray}
      B_z & = & \bar{C}_m e^{(\frac{n}{2}+1)|z|}\left(\chi_m J_\nu
\left(\frac{m}{k}e^{k|z|} \right) + N_\nu \left(\frac{m}{k}e^{k|z|} \right)
\right) \\
B_\mu &=& \partial_\mu \phi
      \label{formula:longitudinal_solution}
\\
      \partial_z \phi& = &\left(1 - \frac{M^2\, a^2(z)}{p^2} \right) B_z
\end{eqnarray}
Here $\bar{C}_m$ is the
normalization constant and
$\nu$ is given by
(\ref{formula:massive_nu}).
The longitudinal modes do not interact with fermions
(this is also true in the extensions of the Standard Model we discuss
in Sections~\ref{3} and~\ref{4}), so we ignore them in what follows.

After integrating
over the coordinate of extra dimension $z$
we obtain the following expression for the effective four-dimensional
action for the transverse massive bulk vector field:
\begin{multline}
	S_{eff}=\int dm\, d^4x \left(-\frac{1}{4}
B_{\mu\nu}(x,m)B_{\mu\nu}(x,m) + \frac{m^2}{2}B_\mu(x,m) B_\mu(x,m) \right)+ \\
	+g_5 \int dm\, \Psi(0,m) \int d^4x\, B_\mu(x,m) \bar{\psi}\gamma^\mu\psi
	\label{formula:effective_action_massive}
\end{multline}

\begin{figure}[!h]
\begin{center}
  \def\svgwidth{0.5\columnwidth}
  %
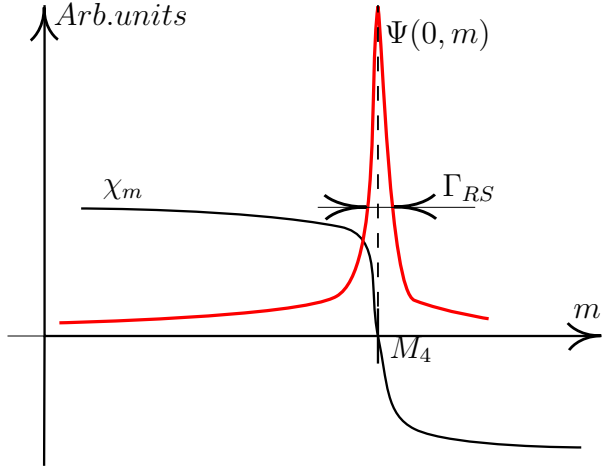%

  \parbox{0.8\columnwidth}{\caption{Quasi-localization of
  massive vector particle: relative weight~$\chi_m$ in~\eqref{eq:aug18-1}
  and wave fiunction at the brane position~$\Psi(m,0)$.
  \label{fig:chi_m} }}
  \end{center}
\end{figure}

In the case of massive bulk vector field, there is no bound state
analogous to \eqref{eqn:photon_solution}, i.e.,
the massive field is not localized on the brane
in the strict sense. Rather, it is quasi-localized on the brane. This effect
is similar to that
studied in Ref~\cite{Dubovsky:2000am} for
massive scalar field. One way to
describe it is to notice that the wave function at the brane
position, and hence the interaction term
in the effective 4-dimensional action (\ref{formula:effective_action_massive})
strongly depends on the 4-dimensional mass,
\begin{equation}
\Psi^2(0,m)=
\frac{\Gamma^2(\frac{n}{2}+1)}{\pi^2(1+\chi^2_m)}
\cdot\left(\frac{m}{2k}\right)^{-n-1}=\frac{n^2}{4}\cdot
\frac{m^4 }{ \Gamma^2(\frac{n}{2}) }\cdot \left(
\frac{m}{2k}\right)^{n-1} \frac{1}{(m^2-M^2_4)^2+m^2
\Gamma_{RS}^{2}},
\label{B0}
\end{equation}
where
\begin{equation}
M_4=M_5 \sqrt{\frac{n}{n+2}} \; , \quad \;\;\;
\Gamma_{RS}=\frac{2\pi}{n\Gamma^2\left(\frac{n}{2}\right)} m
\left(\frac{m}{2k}\right)^n \; .
\label{invisible}
\end{equation}
This is shown in Fig.~\ref{fig:chi_m}. The resonance at $m=M_4$ corresponds precisely to  the
quasi-localized vector particle of 4-dimensional mass $M_4$ and invisible width $\Gamma_{RS} (M_4)$.
Note that in the limit $k\rightarrow\infty$, one has $\Gamma_{RS}\rightarrow 0$, and $ \Psi^2(0,m)$
tends to the delta function:
\begin{equation}
\Psi^2(0,m) \to \frac{nk}{2}\cdot \frac{1}{\pi}
\frac{\Gamma_{RS}}{2}\frac{1}{(m-M_4)^2+\left(\frac{\Gamma_{RS}}{2}\right)^2}\rightarrow
\frac{nk}{2} \cdot \delta(m-M_4). \label{deltafunc}
\end{equation}
This shows that the relation \eqref{eq:aug18-2} between the couplings
remains valid at $M_4 \ll k$. Away from the resonance,
the wave functions at the brane position are still given
by \eqref{A0}.

\subsection{Annihilation into photon and nothing \label{subsubsec:cross-sect}}

\begin{figure}[!h]
  \begin{center}
  \def\svgwidth{0.5\columnwidth}
  %
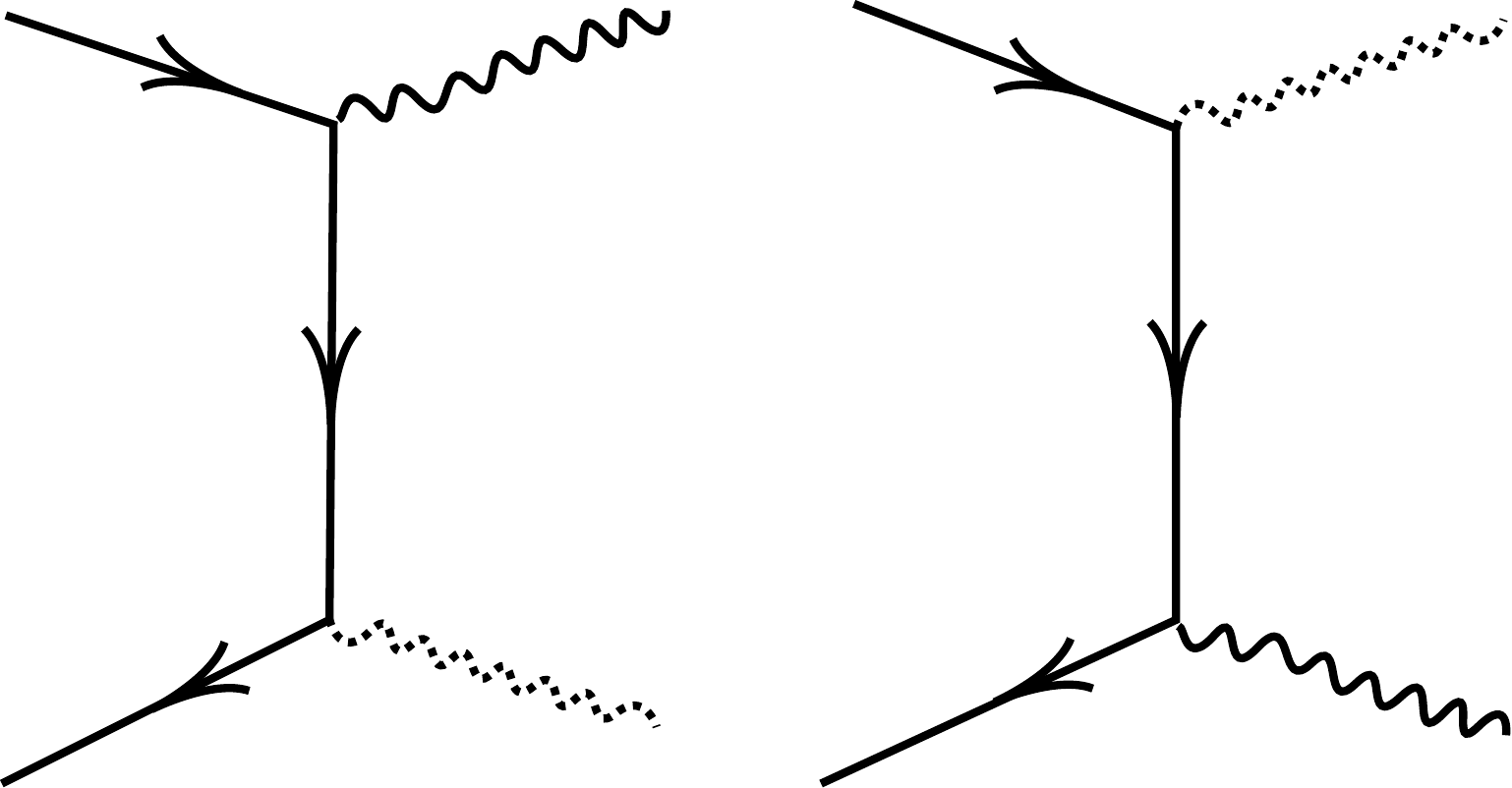%

  \caption{$\bar{e^-} e^+ \rightarrow \gamma + nothing$}
  \label{fig:ff_nothing}
  \end{center}
\end{figure}

Now let us evaluate
the cross-section of fermion-antifermion annihilation
into a photon localized on the brane and invisible vector boson escaping
from the brane. Here we make use of the
effective actions (\ref{formula:effective_action}),
(\ref{formula:effective_action_massive}) for bulk vector bosons, while
the photon is described by the standard electromagnetic action.
The relevant diagrams are shown in Fig.~\ref{fig:ff_nothing}.
The amplitude is given by
\begin{multline}
 i\mathcal{M}=
\overline{v}(p_{2}) (-ig_5 \gamma^{\mu})
\frac{(\hat{p}_{1}-\hat{q})}{(p_{1}-q)^2}
\tilde{e}_{\mu}(\tilde{q})(-ie\gamma^{\nu})u(p_{1})\Psi(0,m)e_{\nu}(q)
+ \\
+ \overline{v}(p_{2}) (-i g_5 \gamma^{\mu})
\frac{(\hat{q}-\hat{p}_{2})}{(q-p_{2})^2}
e_{\mu}({q})(-ie\gamma^{\nu})u(p_{1})B_{m}(0)\tilde{e}_{\nu}(\tilde{q}),
\end{multline}
where $p_{1}$ and $p_{2}$ are the momenta  of the fermion and
antifermion, respectively, $q$ and
$\tilde{q}$ are the momenta of the photon and  vector boson,
$e_\mu$ and $\tilde{e}_\mu$ are their polarization vectors.

We make use of  the expression (\ref{formula:5D_phase_volume}) for the
phase space volume element and obtain the differential
cross section in the rest frame of fermions,
\begin{equation}
    \frac{d\sigma}{dx \, dy}=\frac{g_5^2\,e^2}{8\pi}
\cdot |\Psi(0,M)|^2 \cdot \frac{1}{s} \cdot f(x,y) \; ,
    \label{formula:cross_section_vec_photon}
\end{equation}
where
\begin{equation}
    f(x,y) = \frac{1}{x \sqrt{1-x} } \cdot \frac{(2-x)^2+x^2y^2}{1-y^2} \; .
\label{aug20-2}
\end{equation}
Hereafter
\be
x =2q /\sqrt{s}\; , \quad \;\; y=\cos \theta\; ,
\label{aug20-1}
\ee
where $\theta$ is the angle between the photon
and beam directions, $M=\sqrt{s(1-x)}$
is the 4-dimensional mass of the invisible vector boson.

As an example, let us consider electrodynamics with the bulk
electromagnetic field and brane electrons. This is precisely the theory
of Section~\ref{subsubsec:massless}. We identify $g_4$ entering
\eqref{eq:aug18-2} with $e$, make use of
(\ref{A0}) and obtain
\begin{equation}
    \frac{d\sigma}{dx \, dy} = \frac{4 \alpha^2}{\pi n s} \cdot \frac{1}{\sqrt{1-x}} \cdot \frac{1}{J^2_{\frac{n}{2}}(\frac{M}{k}) + N^2_{\frac{n}{2}}(\frac{M}{k})} \cdot f(x,y)
\end{equation}
In the limit $s \ll k$ we obtain the following expression
\begin{equation}
    \frac{d\sigma}{dx \, dy}=
    \frac{4\pi \alpha^2 }{ n \Gamma^2(\frac{n}{2})}\cdot \frac{1}{s} \cdot
    \left(\frac{\sqrt{s}}{2k}\right)^{n}\cdot (1-x)^{\frac{n}{2}} \cdot
    f(x , y)
    \label{formula:cross_section_vec_photon_u1}
\end{equation}
The cross section rapidly grows with energy for $\sqrt{s} \ll k$, which is a common property
of multi-dimensional theories.

\section{$SU(2)_W\times U(1)_Y$ in the bulk \label{section:full_bulk}}
\label{3}
 Let us now consider $(4+1+n)$-dimensional $SU(2)_{W}\times
U(1)_Y$ gauge theory with bulk gauge fields $A^\alpha_{M}$, $B_M$
and bulk Higgs field
$\Phi$ in the background metric (\ref{gauge_metric}). We still assume
that fermions are localized on the brane.
The action of this theory is
\begin{equation}
S=\int d^4 x\,dz\prod^n_{i=1} \frac{d\theta_i}{2\pi
R_i}\sqrt{g}\left[-\frac{1}{4}\left(F^\alpha_{MN}\right)^2
-\frac{1}{4}B^2_{MN}+\left(D_M \Phi\right)^\dagger D_M\Phi
-V(\Phi^\dagger,\Phi
)+\delta(z)\mathcal{L}_f\right],\label{theactionofthetheory}
\end{equation}
where $\mathcal{L}_f$ is the fermion Lagrangian
(we neglect fermion masses below): 
\begin{multline}
\mathcal{L}_{f}=i\bar{L}\left(\hat{\partial}-
i\widetilde{g}_{1}\frac{Y_{L}}{2}\hat{B}-i\widetilde{g}
\frac{\sigma_{\alpha}}{2}A^{\alpha}\right)L
 +i\bar{r}\left(\hat{\partial}-
i\widetilde{g}_{1}\frac{Y_{R}}{2}\hat{B}\right)r+
 \\+i\bar{Q}\left(\hat{\partial}-i\widetilde{g}_{1}
\frac{Y^Q_{L}}{2}\hat{B}-i\widetilde{g}\frac{\sigma_{\alpha}}{2}A^{\alpha}\right)Q
 +i\bar{q}\left(\hat{\partial}-i\widetilde{g}_{1}
\frac{Y^{q}_{R}}{2}\hat{B}\right)q \; .
\label{ferm gauge int}
\end{multline}
Here $\widetilde{g}$ and $\widetilde{g}_1$ are the
$SU(2)_L\times U(1)_Y$ bulk couplings, $L$ and $r$ denote lepton doublets and singlets respectively, $Q$ and $q$ are analogous quark structures,  and
$V(\Phi^\dagger,\Phi)$ is the Higgs potential,
\begin{equation}
V(\Phi^\dagger,\Phi)
=\frac{\tilde{\lambda}}{2}\left(\Phi^\dagger\Phi-\frac{v^2}{2}\right)^2\; .
\end{equation}
The covariant derivative is, as usual,
$D_M\Phi=\partial_M\Phi-
i\frac{\widetilde{g}_{1}}{2}\hat{B}_M\Phi-i\widetilde{g}
\frac{\sigma_{\alpha}}{2}A^{\alpha}_M\Phi$.
The quadratic action for the vector fields in the Higgs vacuum
is obtained precisely in the same way as in the 4-dimensional Standard Model.
We
make the usual redefinition of the gauge fields,
\[
Z_{M}=\frac{1}{\sqrt{\widetilde{g}^2_{1}
+\widetilde{g}^2}}
\left(-\widetilde{g}_{1}B_{M}+\widetilde{g}A^3_{M}\right),
\quad
A_{M}=\frac{1}{\sqrt{\widetilde{g}^2_{1}+\widetilde{g}^2}}
\left(\widetilde{g}
B_{M}+\widetilde{g}_{1}A^3_{M}\right),
\]
\[W_{M}^\pm=\frac{1}{\sqrt{2}}\left(A_{M}^{1}\mp
iA_{M}^{2}\right)
\]
and write
\begin{equation}
\begin{split}
S&=\int d^4 x\,dz\prod^n_{i=1} \frac{d\theta_i}{2\pi R_i}\sqrt{g}
\Bigl[-\frac{1}{2}|\,W_{MN}\,|^2+m^2_W
|\,W_M\,|^2-\frac{1}{4}Z^2_{MN}+\frac{1}{2}m^2_Z
Z^2_M \\ & \qquad \qquad \qquad \qquad \qquad -\frac{1}{4}F^2_{MN}+
 \delta(z)\mathcal{L}_{f}\Bigr], \label{the action of the
theory broken symm}
\end{split}
\end{equation}
where
\[
m_W^2=\frac{1}{4}\widetilde{g}^2 v^2 \; , \qquad
m_Z^2=\frac{1}{4}(\widetilde{g}^2+\widetilde{g}_{1}^2)v^2
\]
are the bulk masses of the gauge fields, while photon remains massless
in the bulk.
$\mathcal{L}_f$ is the fermion
Lagrangian:
\begin{equation}
\mathcal{L}_f=\mathcal{L}_{f,em}+\mathcal{L}_{f,weak}
\end{equation}
where
\begin{equation}
\mathcal{L}_{f,em}=g_5\sum_f q_f \bar{f}\gamma^\mu A_\mu f,
\end{equation}
and
\begin{equation}
\mathcal{L}_{f,weak} = \frac{\widetilde{g}}{2\sqrt{2}}
\left( \bar{\nu} \gamma^\mu (1-\gamma^5) W^+_\mu  e +h.c.\right)
+\frac{\widetilde{g}}{2\sqrt{2}}
\left( \bar{u} \gamma^\mu (1-\gamma^5) W^+_\mu  d +h.c.\right)
\end{equation}
\[
+\frac{\widetilde{g}}{2 \cos \theta_W}
\sum_f \bar{f} \gamma^\mu
\left( t^f_3 (1-\gamma^5) - 2q_f \sin^2 \theta_W\right) f Z_\mu \; .
\]
Here $
g_5=e \sqrt{\frac{2}{nk}} = \widetilde{g} \sin\theta_W =\widetilde{g}_1
\cos\theta_W $ is the bulk electric charge,
and $eq_f$ is the fermion electric charge. Note that the relation
\eqref{eq:aug18-2} between 4-dimensional and
multi-dimensional couplings is valid for both gauge groups, so $\sin \theta_W$
is expressed in terms of the 4-dimensional gauge couplings in the
standard way.

\subsection{Cross section of $e^+e^- \to \gamma +\mbox{nothing}$}

We see that the model under discussion is a straightforward
modification of the model of Section~\ref{2}.
In particular, the spectrum and the wave functions of the
bulk vector bosons are the same as in
Section \ref{section:wave_functions}.

We consider the process  with the $e^+\,e^-$-pair
in the initial state and with brane photon and bulk $Z$-boson in the
final state.   The  differential cross-section for this process
is
\begin{multline}
    \frac{d \sigma}{dx dy}  \left(e^+ e^- \rightarrow Z_{bulk} \gamma \right) =
    \frac{2 \alpha^2}{ \pi  n k^2}\cdot \frac{(-\frac{1}{2} + \sin^2 \theta_W )^2 + \sin^4 \theta_W}{\sin^2 \theta_W \cos^2 \theta_W} \cdot  \\
    \cdot \frac{\sqrt{1-x}}{\left((\frac{n}{2} + 1 - \nu) J_\nu(\frac{M}{k}) + \frac{M}{k} J_{\nu - 1} (\frac{M}{k}) \right)^2+
    \left((\frac{n}{2} + 1 - \nu) N_\nu(\frac{M}{k}) + \frac{M}{k} N_{\nu - 1} (\frac{M}{k}) \right)^2} \cdot f(x,y)
\end{multline}
\[
    \nu=\sqrt{\left(\frac{n}{2} + 1\right)^2 + \frac{n + 2}{2} \left(\frac{M_Z}{k}\right)^2}, \quad M=\sqrt{s (1 - x)}
\]
In the limit $s \ll k$ we obtain

\begin{multline}
\frac{d \sigma}{dx dy}\left(f\bar{f}\rightarrow Z_{bulk}
\gamma \right) = \frac{\pi \alpha^2}{s\; n \Gamma^2\left(\frac{n}{2}\right)} \cdot \frac{8 \sin^4\theta_W- 4\sin^2\theta_W + 1}{\sin^2\theta_W \cos^2\theta_W} \cdot
\left(\frac{\sqrt{s}}{2k}\right)^n  \cdot
 \\ \cdot \frac{(1-x)^2}{(1-x-\frac{M^2_Z}{s})^2+
\frac{M^2_Z \Gamma_{RS}^2}{s^2}} \cdot f(x,y) \; .
\label{finalRS2crosssection}
\end{multline}
Here $M_Z$ is the four-dimensional $Z$-boson mass, 
and $\Gamma_{RS}$ is given by (\ref{invisible}). 
Other notations are the same as
in \eqref{aug20-2}, \eqref{aug20-1}.
The $x$-dependent
factor in the last part of
the formula (\ref{finalRS2crosssection}) comes from
the squared wave function at the brane position, see (\ref{B0}).

The total cross section of
 $e^+e^- \to \gamma +\mbox{nothing}$ is the sum
\begin{equation}
\frac{d \sigma}{dx dy}\left(f\bar{f}\rightarrow
\gamma +\mbox{nothing}\right)
= \frac{d \sigma}{dx dy}\left(f\bar{f}\rightarrow Z_{bulk}
\gamma \right) + \frac{d \sigma}{dx dy}\left(f\bar{f}\rightarrow
\gamma_{bulk}
\gamma \right) \; ,
\label{formula:cross-set-full-bulk}
\end{equation}
where the first term is given by \eqref{finalRS2crosssection} and the second
term is the cross section of the production of bulk photon which is
given by \eqref{formula:cross_section_vec_photon_u1}.

\section{$SU(2)_{brane} \times U(1)_{bulk}$ \label{section:half_bulk}}
\label{4}

In this Section we consider a model with localized
fermions, the Higgs field
localized on the brane and  $SU(2)_{brane} \times U(1)_{bulk} $
gauge group. So, among all fields of the Standard Model,
only $U(1)_Y$ gauge field propagates in extra dimensions.
An interesting feature of this model is that after 
spontaneous symmetry breaking, the set of fields consists of the localized on 
the brane photon $A_\mu$ and quasi-localized $B_\mu$ instead of $Z_\mu$. 
Nevertheless, the Standard Model 
is restored in the limit $k \rightarrow \infty$.

Let us begin with
the effective action for this theory. After integrating
out compact extra dimensions, the action for this model is written as follows:
\begin{equation}
    S=\int d^4 x dz a^{n+4} \left( \left( (D_\mu \phi)^+ D_\mu \phi  -
\frac{1}{4}F_{\mu\nu}^\alpha F_{\mu\nu}^\alpha -
\lambda(\phi^{+}\phi-\frac{v^2}{2}) \right)\delta(z) - \frac{1}{4}
g^{MA} g^{NB} B_{MN}B_{AB}  \right)
    \label{eqn:action_su2_u1}
\end{equation}
Here $B_{M}(x,z)$ is the $U(1)_Y$ gauge field,
$A^\alpha_\mu(x)$, $\alpha=1,2,3$  are the $SU(2)_{brane}$ gauge fields,
$F^\alpha_{\mu\nu}=\partial_\mu A^\alpha_\nu - \partial_\nu
A_\mu^\alpha + g\varepsilon^{\alpha bc}A^b_\mu A^c_\nu$,
$D_\mu \phi = \partial_\mu \phi -
\frac{g}{2}\tau_\alpha A_\mu^\alpha -
\frac{g^\prime}{2}\sqrt{\frac{2}{kn}} B_\mu(x,0)$.
When writing the last expression we have recalled
the relationship \eqref{eq:aug18-2} between
the multi-dimensional and
4-dimensional couplings.

After spontaneous symmetry breaking, the linearized field
equations for electrically neutral vector fields are
\begin{equation}
  \partial_z \partial_\mu B_\mu  =  0
    \label{eqn:B_gauss}
\end{equation}
\begin{equation}
     \frac{1}{a^2} \partial_\mu \partial_\mu  B_{\nu} -   \partial_z^2 B_\nu + (n+2)ksign(z) \partial_z B_\nu = \frac{2v^2 g^{\prime}}{8}\sqrt{\frac{2}{kn}} \left(g^{\prime}\sqrt{\frac{2}{kn}}B_\nu - g A^3_\nu \right)\delta(z)
    \label{eqn:B_z}
\end{equation}
\begin{equation}
    \partial_\mu F_{\mu\nu} + \frac{2gv^2}{8}\left( g A_\nu^3 - g^{\prime}\sqrt{\frac{2}{kn}} B_\nu(x,0) \right)=0
    \label{eqn:A_u1}
\end{equation}
Here we use the gauge $B_z = 0$. The equations for $W^\pm$ bosons have the standard four-dimensional form, so we do not consider them.
Equations (\ref{eqn:B_z}), (\ref{eqn:A_u1}) have a peculiarity: since the Higgs mechanism operates on the brane, there is an additional boundary condition on the brane.

Let us first consider the case  $A^3_\mu(x) = \frac{g^\prime}{g}\sqrt{\frac{2}{kn}} B_\nu(x,0)$. Equations (\ref{eqn:B_gauss})-(\ref{eqn:A_u1}) take the following form
\begin{eqnarray}
\partial_\mu F_{\mu\nu}=0 \label{eqn:photon_a3}\\
g^{\prime}B^\prime_\nu - g A^3_\nu  =0 \label{eqn:photon_B1} \\
\frac{1}{a^2} \partial_\mu \partial_\mu  B_{\nu} -   \partial_z^2 B_\nu + (n+2)ksign(z) \partial_z B_\nu = 0 \\
\partial_\mu \partial_z B_\mu = 0  \label{eqn:photon_B2}
\end{eqnarray}
Equation (\ref{eqn:photon_a3}) has a solution with non-zero mass only if it is
longitudinal, $A^3_\mu=p_\mu b(p)$, but such a solution
is incompatible with Eqs. (\ref{eqn:photon_B1})-(\ref{eqn:photon_B2})
for $p^2 \neq 0$. Thus, Eqs.(\ref{eqn:photon_a3})-(\ref{eqn:photon_B2}) describe a massless particle.
We have shown in Section~\ref{section:wave_functions} that massless
solution to equations (\ref{eqn:photon_B1}),(\ref{eqn:photon_B2})
is localized on the brane.
So Eqs. (\ref{eqn:photon_a3})-(\ref{eqn:photon_B2})
describe the photon field $A_\mu$. The fields $B_\mu$, $A^3_\mu$ are related to $A_\mu$ as follows
\begin{equation}
 B_\mu(p,z) = \sqrt{\tfrac{kn}{2}} \tfrac{g}{\sqrt{{g^\prime}^2+g^2}}\cdot A_\mu(p)   
\end{equation}
\begin{equation}
A^3_\mu(p)  =  \tfrac{g^\prime}{\sqrt{{g^\prime}^2+g^2}}\cdot A_\mu(p)
\end{equation}

The other case is $A^3_\mu(x) \neq \frac{g^\prime}{g}\sqrt{\frac{2}{kn}}
 B_\nu(x,0)$. The situation here is more complicated
than the previous one.
Equation (\ref{eqn:B_z}) can be rewritten as bulk equation and boundary condition on the brane:
\begin{eqnarray}
     \frac{1}{a^2} \partial_\mu \partial_\mu  B_{\nu} -   \partial_z^2 B_\nu + (n+2)ksign(z) \partial_z B_\nu = 0 \label{eqn:B_massive_z}
      \\
     \left.\partial_z B_\nu \right|_{z\rightarrow +0} - \left.\partial_z B_\nu\right|_{z\rightarrow -0} = \frac{2g^\prime v^2}{8}\sqrt{\frac{2}{kn}}\left(gA_\nu^3-g^\prime\sqrt{\frac{2}{kn}} B_\nu \right)
     \label{eqn:B_boundary_condition}
\end{eqnarray}
Eqs.~(\ref{eqn:B_massive_z}),~(\ref{eqn:B_gauss}) are the same
as in the case of the bulk-massless vector field that was discussed in
Section~\ref{section:wave_functions}. Therefore we use the
solution~(\ref{gauge_solution}),~(\ref{formula:vertex_prefactor})
for the field $B_\mu(x,z)$, but the presence of the boundary
condition~(\ref{eqn:B_boundary_condition}) leads to the different
expression for~$\eta_m$. $\eta_m$ and $A^3_\mu$ are obtained from
Eqs.~(\ref{eqn:A_u1}),~(\ref{eqn:B_boundary_condition}):
\begin{equation}
  \eta_m = - \frac{ - \frac{2m {g^\prime}^2 }{kn} N_\nu(\frac{m}{k}) + 2 \left( \frac{4 m^2}{v^2} - g^2 \right) N_{\nu -1}(\frac{m}{k}) }{ - \frac{2m {g^\prime}^2 }{kn} J_\nu(\frac{m}{k}) + 2 \left( \frac{4 m^2}{v^2} - g^2 \right) J_{\nu -1}(\frac{m}{k}) }, \quad \nu = \frac{n}{2} + 1
  \label{formula:eta_m_half_bulk}
\end{equation}
\begin{equation}
  A_\mu^3(p,m)= - B_\mu(p,m)\Psi(0,m)\frac{g g^\prime v^2 \sqrt{\frac{2}{kn}}}{4m^2 - v^2g^2 }
  \label{eqn:A3_solution}
\end{equation}
where $\Psi(0,m)$ and $B_\mu(p,m)$ have the same form as in 
(\ref{formula:vertex_prefactor}) and (\ref{gauge_solution}) respectively 
with $\eta_m$ given by (\ref{formula:eta_m_half_bulk}).
The normalization constant for this solution has been straightforwardly 
obtained by making use of the approach outlined in 
Section~\ref{subsubsec:massless}. 
It is worth noting that $\Psi(0,m)$ has the following explicit form
\begin{equation}
    \Psi(0,m)=(m^2 - M^2_W) \widetilde{\Psi}(0,m)
\end{equation}
\begin{equation}
    \widetilde{\Psi}(0,m)=\sqrt{\frac{m}{2k}}
      \frac{2\left(J_{\nu-1}(\tfrac{m}{k}) N_{\nu}(\tfrac{m}{k})-N_{\nu-1}(\tfrac{m}{k}) J_{\nu}(\tfrac{m}{k})\right)}
           {\sqrt{\left( (m^2-M_W^2)J_{\nu-1}(\tfrac{m}{k}) - \tfrac{m}{kn}\tg^2\theta_W M^2_W J_\nu(\tfrac{m}{k}) \right)^2 + (J_\nu \rightarrow N_\nu)^2}}\, .
    \label{formula:Psi_half_bulk}
\end{equation}
Thus, there is no pole in (\ref{eqn:A3_solution}) at $m=M_W=\tfrac{g^2v^2}{4}\,$.

In order to illustrate how the the massive vector fields are
quasi-localized in this theory, we write the following 
expressions for the fields $A_\mu$, $Z_\mu$ 
at the brane position,
\begin{equation}
  Z_\mu(p,m)=\tfrac{1}{\sqrt{g^2+{g^\prime}^2}}\cdot \left(gA^3_\mu(p,m)- g^\prime\sqrt{\tfrac{2}{kn}}B_\mu(p,m)\right)=m^2 \sin{\theta_W}\cdot B_\mu(p,m) \widetilde{\Psi}(0,m)  
  \label{formula:bulk_Z}
\end{equation}
\begin{equation}
  A_\mu(p,m)=\tfrac{1}{\sqrt{g^2+{g^\prime}^2}} \left(g^\prime A^3_\mu(p,m)+g \sqrt{\tfrac{2}{kn}}B_\mu(p,m)\right)=(m^2 - M_Z^2) \cos{\theta_W}\cdot B_\mu(p,m)  \widetilde{\Psi}(0,m)  
  \label{fomula:bulk_photon1}
\end{equation}
\begin{figure}[!h]
    \begin{center}
      \def\svgwidth{0.6\columnwidth}
      %
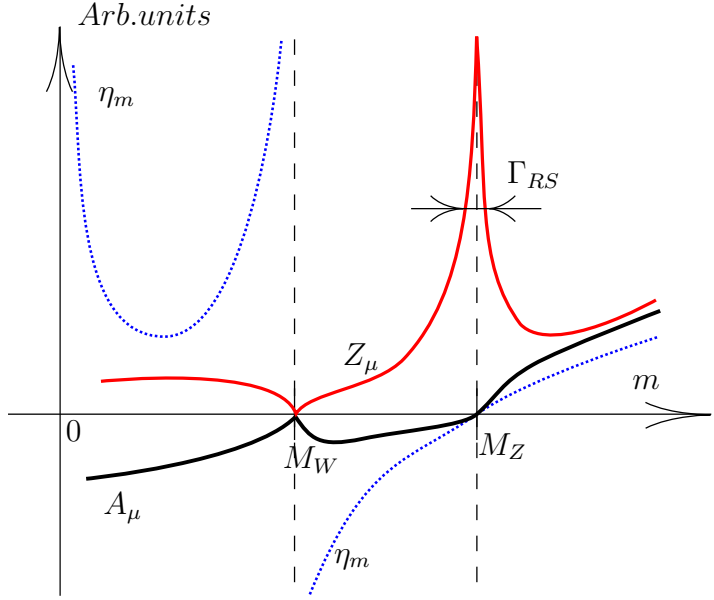%

      \caption{$A_\mu$, $Z_\mu$, $\eta_m$ as functions of $m$}
      \label{fig:A_Z_eta}
    \end{center}
\end{figure}
It is worth noting that $Z_\mu(p,m)$ as function of $m$, which is plotted in
Fig.\ref{fig:A_Z_eta}, has a peak at $m=M_Z$ and 
the position of this peak corresponds to the zero of~$A_\mu(p)$.
The width $\Gamma_{RS}$ of this peak is obtained from 
Eqs.~(\ref{formula:Psi_half_bulk}),~(\ref{formula:bulk_Z}) 
and is given by
\begin{equation}
    \Gamma_{RS}(M_Z) = \frac{2 \pi M_Z \sin^2 \theta_W}{n \Gamma^2(\tfrac{n}{2})}\left(\frac{M_Z}{2k} \right)^n
    \label{formula:Gamma_RS}
\end{equation}
Since the fields $A_\mu(p,m)$ and $Z_\mu(p,m)$ are not independent,
we can rewrite the interaction terms in the action in terms 
of the independent field $B_\mu$,
\begin{multline}
    S_{int}=\int dm\, d^4x\left( \frac{g(-\tfrac{1}{2}+\sin^2{\theta_W})}{\cos{\theta_W}}Z_\mu \bar{e}_L\gamma^\mu e_L + \frac{g\sin^2{\theta_W}}{\cos{\theta_W}} Z_\mu \bar{e}_R\gamma^\mu e_R -e A_\mu \bar{e}  \gamma^\mu e \right)=  \\
    =\int dm\, d^4x \frac{e \,\widetilde{\Psi}(0,m)}{\cos \theta_W}\cdot \left( \left(\frac{m^2}{2}- M_W^2\right) \bar{e}_L\gamma^\mu e_L 
    +(m^2-M^2_W)  \bar{e}_R\gamma^\mu e_R  \right) B_\mu(x,m) \, ,
\end{multline}
where we consider electrons only. In the limit $k \rightarrow \infty$, 
one has $\Gamma_{RS} \rightarrow 0$ and
\begin{equation}
\widetilde{\Psi}^2(0,m)~\rightarrow~\frac{nk}{2M^2_Z}\delta(m-M_Z)\,.
\end{equation}
Thus, only the localized mode of the field~$Z_\mu$ with mass~$M_Z$
interacts with fermions, while all other massive modes of $Z_\mu$
and $A_\mu$ do not interact with fermions directly. In this way, the four-dimensional physics is restored in the limit $k \rightarrow \infty$ in this model.

Thus, we obtain that the effective vector fields  in this model are  
localized $W^\pm$ bosons, localized photon $A_\mu$, and quasi-localized 
vector field $B_\mu$. Only $B_\mu$ can escape from the brane.

Let us turn to the cross-section of electron-positron 
annihilation to photon and $B_\mu$. 
Its general form is given by (\ref{formula:cross_section_vec_photon}). 
In the center-of-mass frame we have, explicitly,
\begin{multline}
    \frac{d\sigma}{dx\,dy}=\frac{2 \alpha ^ 2 }{\pi n s \cos^2\theta_W}\frac{1}{\sqrt{1-x}}\cdot f(x,y)\cdot
    \\
    \cdot \frac{(m^2-M_W^2)^2 + (\frac{m^2}{2} - M_W^2)^2 }{
            ((m^2-M_W^2) J_{\nu-1}(\frac{m}{k})- \frac{m}{n k} \tg^2\theta_W M_W^2 J_\nu(\frac{m}{k}))^2 + (J_\nu \rightarrow N_\nu)^2
    }
    \label{formula:cross-sec-half-bulk}
\end{multline}
Here $m = \sqrt{s(1 - x)}$. In the limit $s \ll k$  we obtain
\begin{multline}
    \frac{d\sigma}{dx \, dy}=\frac{4\pi \alpha^2 }{ n \Gamma^2(\frac{n}{2})} \cdot \frac{(m^2-M_W^2)^2+\left(\frac{m^2}{2} -M_W^2\right)^2 }{2\cos^2{\theta_W }}\cdot \frac{1}{s}
    \left(\frac{\sqrt{s}}{2k}\right)^{n}\cdot f(x , y) \cdot \\ \cdot \frac{1}{(m^2-M_Z^2)^2+m^2\Gamma_{RS}^2(m^2)}
\end{multline}
where $x$, $y$ and $f(x,y)$ are the same as in  Section~\ref{subsubsec:cross-sect}, and  $\Gamma_{RS}(m^2)$ is given by (\ref{formula:Gamma_RS}).

\section{Signal at $e^+ e^-$--collider  \label{section:signals}}
When discussing the implications of the above results for future
$e^+ e^-$--collider, we have to take into consideration the constraints that already exist, the Standard Model  background and the predictions of competing ADD model. Let us now turn to these issues.

\subsection{Invisible $Z$-decay}
The models we study in this paper have two new parameters, $k$ and $n$.
Strong constraints on the parameter $k$ are obtained by considering
the invisible decay of $Z$-boson. In addition to the Standard Model
invisible decay channels, there is the escape of $Z$-boson
from the brane.
In the $SU(2)_{bulk} \times U(1)_{bulk}$ model, the partial width
of the latter is given by  (\ref{invisible}), i.e.,
\begin{equation}
    \Gamma^Z_{RS} =
M_Z \frac{2\pi}{n\Gamma^2(\frac{n}{2})} \cdot \left(\frac{M_Z}{2k}\right)^n
\; .
\end{equation}
In the model $SU(2)_{brane} \times U(1)_{bulk}$ the invisible decay
width is given by~(\ref{formula:Gamma_RS}) and contains
the additional  factor $\sin^2 \theta_W$.

The measured invisible $Z$-boson decay width
agrees with the Standard Model within the experimental
uncertainty
$\Delta \Gamma^Z_{invis} \simeq 1.5~\mbox{MeV}$~\cite{Amsler:2008zzb}.
We require that the additional invisible decay width does not
exceed this uncertainty and obtain the bounds given in
Table \ref{table:constraints}
\begin{table}
    \begin{center}
    \begin{tabular}{l c  c}
    \toprule
    n   & k (GeV), $SU(2)_{bulk} \times U(1)_{bulk}$ & k (GeV), $SU(2)_{brane} \times U(1)_{bulk}$ \\
    \midrule
    $1$ & $5.5 \cdot 10^6$ & $1.3 \cdot 10^6$ \\
    $2$ & $20 \cdot 10^3$  & $10 \cdot 10^3$  \\
    $3$ & $2.5 \cdot 10^3$ & $1.6 \cdot 10^3$\\
    $4$ & $900$ & $600$ \\
    $5$ & $400$ & $300$ \\
    $6$ & $300$ & $200$ \\
    \bottomrule
    \end{tabular}
  \parbox{0.7\textwidth}{  
    \caption{Constraints on the parameter $k$ from invisible $Z$ boson decay 
             in models of Section~\ref{section:full_bulk} (left) 
             and Section~\ref{section:half_bulk} (right)}
    \label{table:constraints} }
    \end{center}
\end{table}
For models with $n=1$ and $n=2$ and, to lesser extent, $n=3$
these constraints are so strong that the detection of the process
studied in this paper appears hopeless.
So, in what follows we present the results for $n > 3$.

\subsection{ADD cross section for invisible graviton production}

New processes with invisible particles in the
final state are predicted by various
models. For comparison we recall here
 the ADD cross section of $e^+ e^-$ annihilation into photon and
invisible graviton computed in  Ref.~\cite{Giudice:1998ck},
\begin{equation}
\frac{d^2\sigma}{dx d\cos \theta}(e^+e^- \to \gamma G)
=\frac{\alpha}{64}\cdot\frac{2\pi^{n/2}}{\Gamma(n/2)}\cdot
\left( \frac{\sqrt{s}}{M_D}\right)^{n +2}
\cdot \frac{1}{s} \cdot f_{ADD}(x, y) \; ,
\label{formula:ADD_cross_section}
\end{equation}
where
\begin{equation}
f_{ADD}(x,y) = \frac{2(1-x)^{\frac{n}{2} -1}}{x(1-y^2)}
\left[ (2-x)^2(1-x+x^2)-3y^2x^2(1-x)-y^4x^4 \right] \; ,
\end{equation}
$n$ is the number of large extra dimensions and $M_D$ is the
corresponding $n$ - dimensional Planck mass,
which is a free parameter of the ADD model

\subsection{Annihilation into photon and neutrinos}

The Standard Model background is predominantly due to
the process $f  \bar{f}  \rightarrow \gamma \nu \bar{\nu}$.
The $Z$-peak contribution
from $f \bar{f} \rightarrow \gamma Z$ can be eliminated at
$e^+ e^-$-collider by excluding the photon
energy region around  $q = (s -M_Z^2)/(2\sqrt{s})$. The remaining
background due to
$f \bar{f} \rightarrow \gamma \nu \bar{\nu}$
is continuously distributed in $q$.
Other background contributions, e.g., due to
 $e^+e^- \rightarrow \gamma (e^+e^-)$ or
$e^+e^- \rightarrow \gamma (\gamma)$, should not be
important at  large photon transverse energies.

\begin{figure}[h!]
   \begin{center}
      \def\svgwidth{0.6\columnwidth}
      %
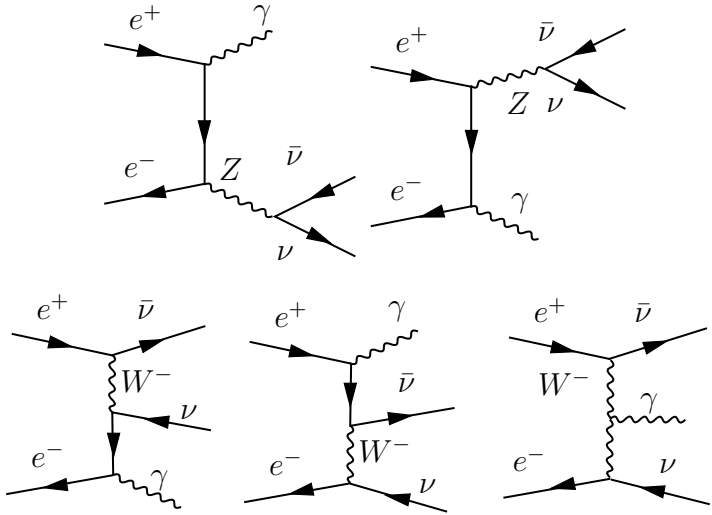%

   \end{center}
    \caption{Tree-level SM background}
    \label{fig:SM_diagrams}
\end{figure}

The cross-section of $e^+ e^- \rightarrow  \gamma \nu \bar{\nu}$ is
due to five diagrams shown in Fig.\ref{fig:SM_diagrams}.
We have computed this cross-section by making use of COMPHEP package \cite{comphep}.

\subsection{$e^+ e^-$ collider missing energy signal}
\label{5}
Let us present the predictions of the models with
$SU(2)_{bulk} \times U(1)_{bulk}$ and $SU(2)_{brane}~\times~U(1)_{bulk}$
which are given by (\ref{formula:cross-set-full-bulk}) and (\ref{formula:cross-sec-half-bulk}) respectively,
together with the Standard Model background
and the prediction (\ref{formula:ADD_cross_section}) of the ADD model.

\begin{figure}
\begin{center}
  \includegraphics[width=0.8\textwidth]{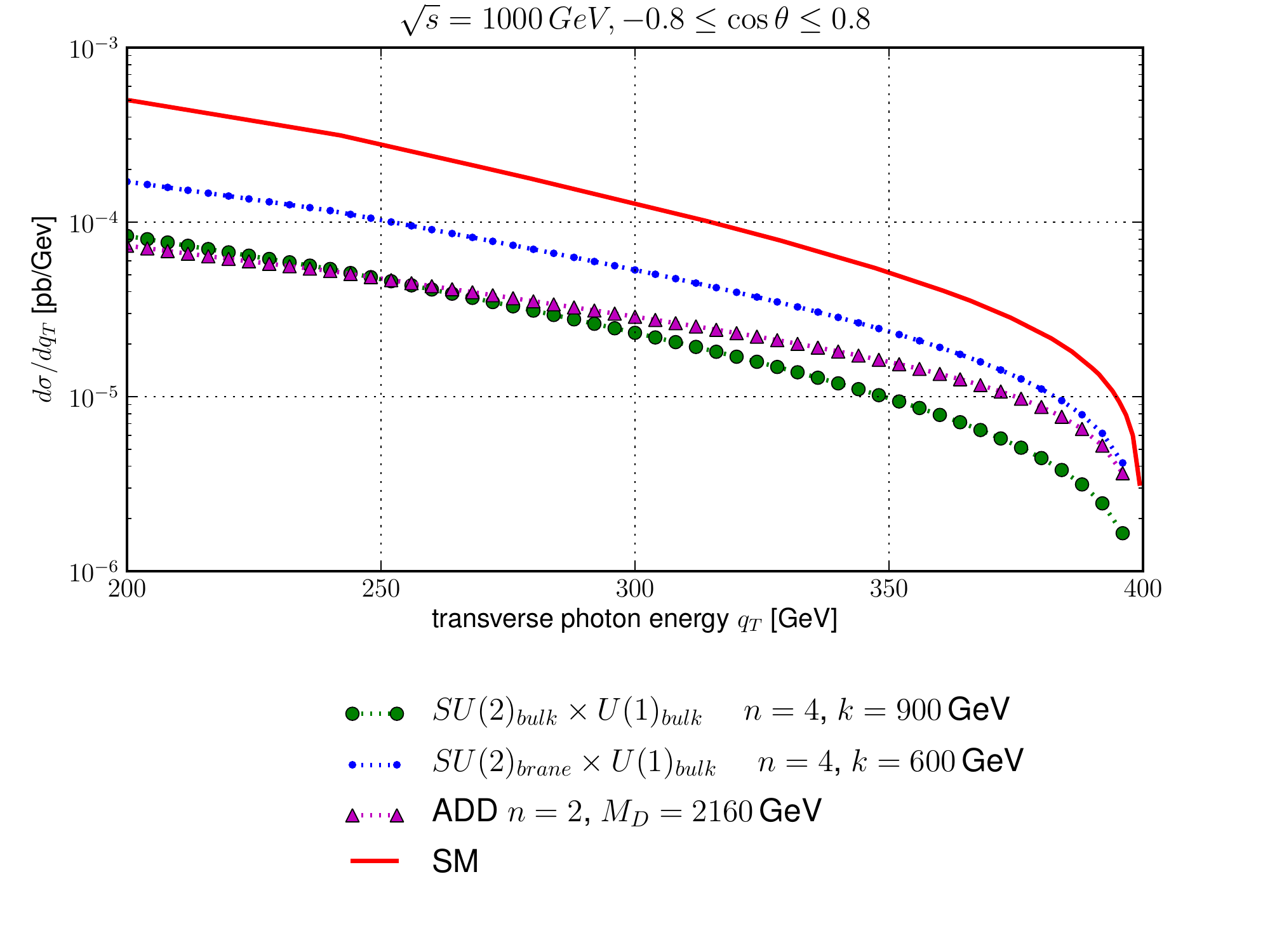}  
  \includegraphics[width=0.8\textwidth]{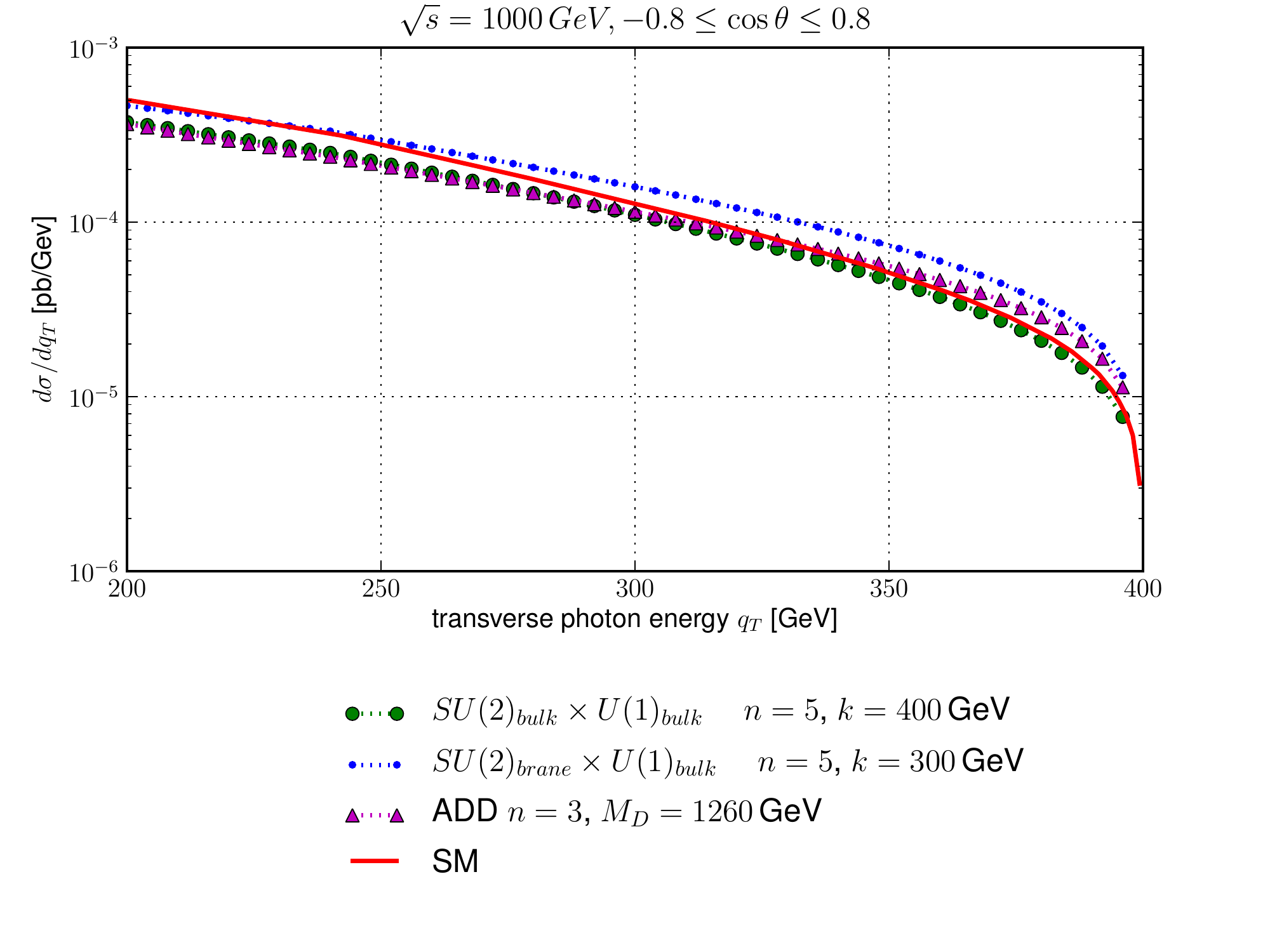}  \\
  \end{center}
  {\caption \qquad
  Differential $e^+e^-\rightarrow \gamma+ \mbox{nothing}$
  cross-section at $\sqrt{s}=1$TeV integrated
  over angle between photon and beam in the range~$-0.8<\cos\theta<0.8$.
  We have imposed the cut $q < 400$ GeV for both signal and
  background. \label{fig:1_TeV_qT} }
\end{figure}

In Fig.\ref{fig:1_TeV_qT} we show the differential cross-section
of $e^+ e^- \rightarrow \gamma +\mbox{nothing}$ at $\sqrt{s}=1$TeV,
integrated over angle $\theta$ in the range $-0.8 < \cos \theta < 0.8$
as function of the transverse photon momentum $q_T$.
The value of the RS parameter $k$ is the lowest one
compatible with the constraints given in the
Table.\ref{table:constraints}. The value of the ADD parameter $M_D$ is
chosen in such a way that the ADD signal and RS signal are the
same when integrated over $-0.8 < \cos \theta < 0.8$ and
$200 \mbox{GeV} <q<400 \mbox{GeV}$.
The cut $q < 400$ GeV is imposed to eliminate the background
contribution from $Z$ peak. We see from Fig.\ref{fig:1_TeV_qT} that the effect we study can be considerable, especially for $n \geq 5$, even in view of the strong constraints coming from the invisible $Z$-decay. On the other hand, the shape of the $q_T$-distribution is not dramatically different from that of the Standard Model background or ADD prediction.

\begin{figure}[p]
  \begin{center}
  \includegraphics[width=0.8\textwidth]{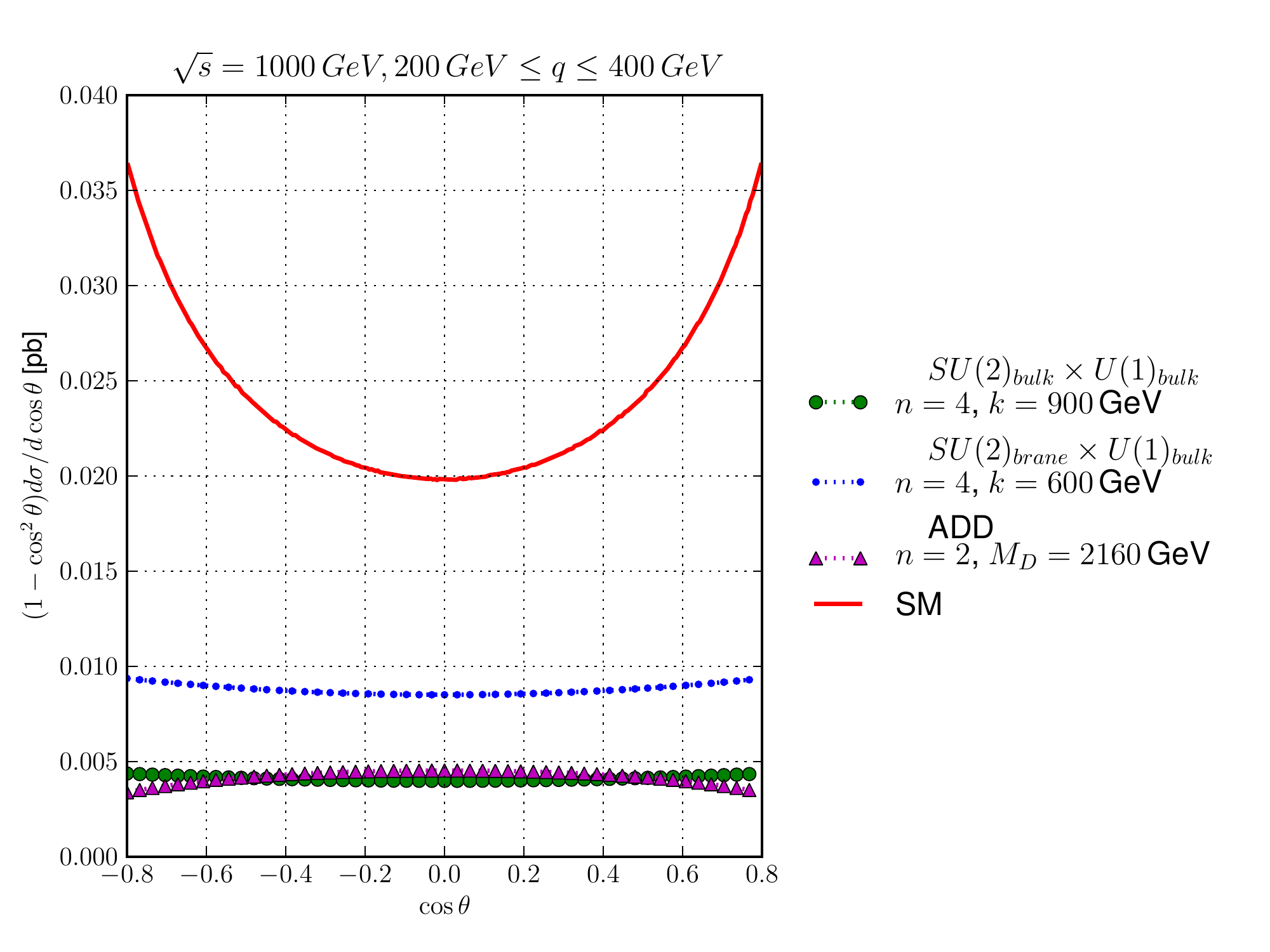}  
  \includegraphics[width=0.8\textwidth]{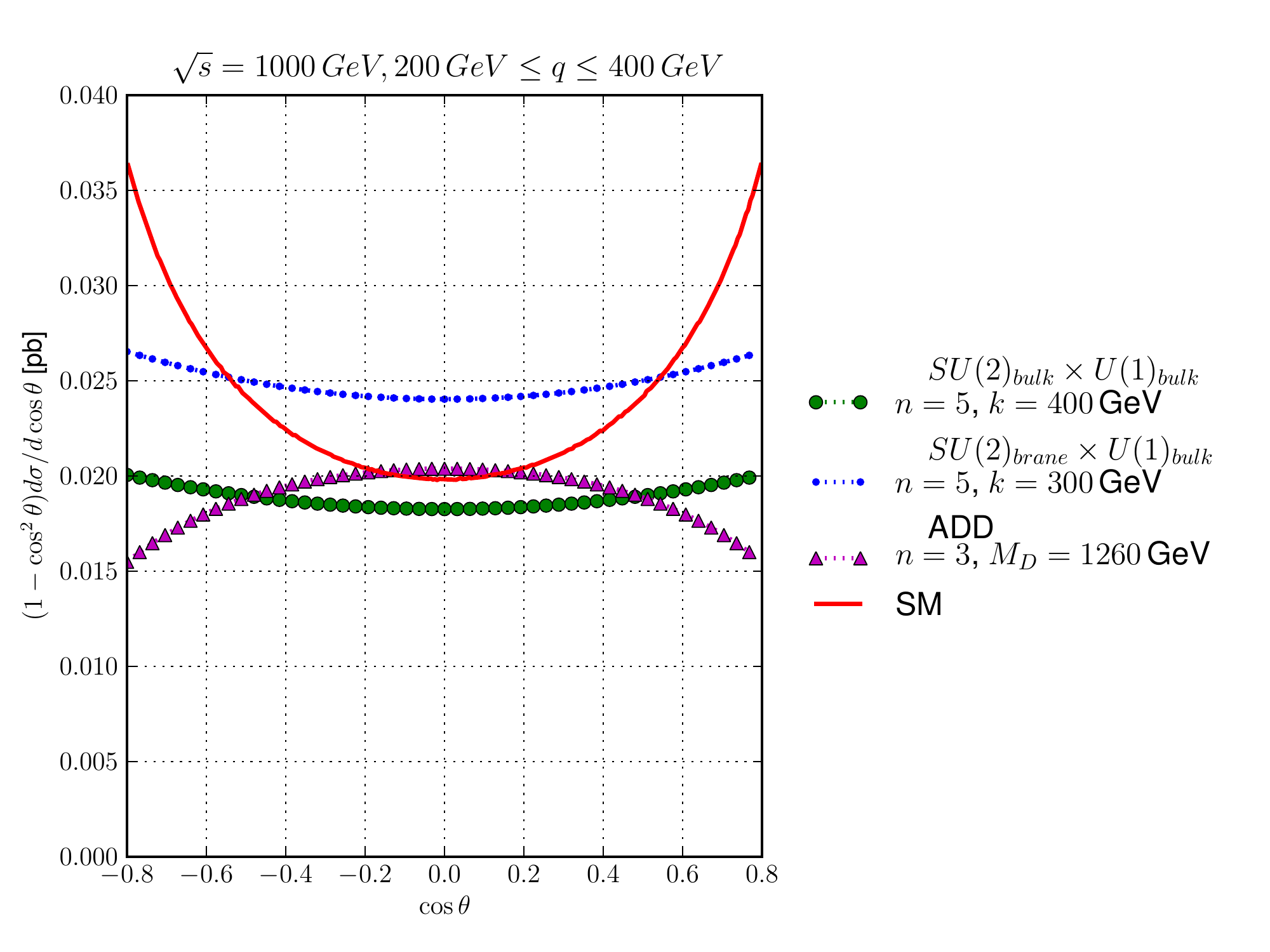} 
  \end{center}
  {\caption \qquad Differential $e^+ e^- \rightarrow \gamma+$nothing cross-section at $e^+e^-$ collider at $\sqrt{s}=1$TeV integrated over photon momentum in the range $200$GeV$< q < 400$GeV. 
  \label{fig:1_TeV_cos} }
\end{figure}

Fig.\ref{fig:1_TeV_cos} shows the differential cross-section
multiplied by $(1 - \cos ^2 \theta )$ for $\sqrt{s}=1$TeV,
integrated over photon momentum in the
range $200 \mbox{GeV} <q<400 \mbox{GeV}$, as function
of $\cos \theta $. 
From Fig.\ref{fig:1_TeV_cos} we conclude that our prediction can be
distinguished from that of the ADD model and from the SM background through the
analysis of the angular distribution of photons with high transverse
momentum, especially for $n \geq 5$. 

The overall conclusion is that our models with $n \geq 5$ will be possible to probe in $e^+ e^- $-annihilation, while the cases $n \leq 4$ are difficult.

\section{Conclusion \label{section:conclusion}}
In this paper we have studied phenomenology of two Standard Model 
extensions in the modified Randall Sundrum II background metric.
We found that the parameters of the models are constrained by the 
measurements of the invisible width of the $Z$ boson.
We computed the cross-section of process 
$e^+e^- \rightarrow \gamma + nothing$ in these models, and found that the signal may be sizeable for $n \geq 5$.

An obvious extension of our analysis would be the study of the processes
with missing energy in the final state at LHC, which are due to the
vector bosons escaping from our brane. We hope to turn to this study 
in future.

We are indebted to V.~A.~Rubakov for helpful discussions and 
suggestions. We also thank S.~N.~Gninenko, D.~S.~Gorbunov, D.~G.~Levkov,
M.~V.~Libanov, and E.~Y.~Nugaev for helpful discussions.
This work was supported in part by the grant of the President of the
Russian Federation NS-5525.2010.2, by the Ministry of Science and
Education under state contracts 02.740.11.0244 and P2598,
grant of the President of the Russian Federation MK-7748.2010.2.


\end{document}